\documentclass[aps,prc,groupedaddress,floatfix]{revtex4}

\usepackage{amsfonts}
\usepackage{graphicx}
\usepackage{verbatim}

\begin{document}


\title{Low-energy QCD}


\author{Marco Frasca}
\email[]{marcofrasca@mclink.it}
\affiliation{Via Erasmo Gattamelata, 3 \\ 00176 Roma (Italy)}


\date{\today}

\begin{abstract}
We derive a low-energy quantum field theory from quantum chromodynamics (QCD) that holds in the limit of a very large coupling. All the parameters of the bare theory are fixed through QCD. Low-energy limit is obtained through a mapping theorem between massless quartic scalar field theory and Yang-Mills theory. One gets a Yukawa theory that, in the same limit of strong coupling, reduces to a Nambu-Jona-Lasinio model with a current-current coupling with scalar-like excitations arising from Yang-Mills degrees of freedom. A current-current expansion in the strong coupling limit yields a fully integrated generating functional that, neglecting quark-quark current coupling, describes all processes involving glue excitations and quark. Some processes are analyzed and we are able to show consistency of Narison-Veneziano sum rules. Width of the $\sigma$ resonance is computed. The decay $\eta'\rightarrow\eta+\pi^++\pi^-$ is discussed in this approximation and analyzed through the more elementary processes $\eta'\rightarrow\eta+\sigma$ and $\sigma\rightarrow\pi^++\pi^-$. In this way we get an estimation of the mass of the $\sigma$ resonance and the value of the $\eta$ decay constant. This $\eta'$ decay appears a possible source of study for the $\sigma$ resonance.
\end{abstract}


\maketitle


\section{Introduction}

With the discovery of asymptotic freedom \cite{gross,poli} and the settlement of a theory of strong interactions with quantum chromodynamics (QCD), the question of the determination of the low-energy behavior of the theory has been increasingly demanding. Once it was clear that the high-energy behavior of strong interactions was properly described, the emergence of the low-energy spectrum of the theory must have been obtained. This question encountered a serious difficulty as perturbation theory is useless in this case due to the strength of the coupling and there were no useful mathematical techniques at that time. At the very start of this story, Wilson come out with a discretized approach for QCD \cite{wils}. This approach made the basis for the birth of lattice QCD that still today is the most reliable approach to solve QCD at low energies.

In the course of time, some approaches have been devised to manage low-energy QCD. The idea behind this is to move to a phenomenological theory having all the properties of QCD. With this view in mind, a very successful approach was devised by Weinberg \cite{wei}, Gasser and Leutwyler \cite{gleu}: Chiral perturbation theory (ChPT). Since then, this approach has found increasing success (for a review see \cite{sche}) and experimental support \cite{na48}. This approach introduces some phenomenological constants that are fixed through experimental measurements. Analyses of weak hadronic decays are accomplished through this approach \cite{sre}.

On a different side, Nambu-Jona-Lasinio model \cite{njl1,njl2} proved to be a very satisfactory description of hadronic phenomenology \cite{kle}. In this case the model is based on a quark current-current interaction and is phenomenologically built to display dynamical chiral symmetry breaking. In its original formulation, Nambu-Jona-Lasinio model is neither renormalizable nor confining. A proposal to modify this situation has been given in \cite{rkl} and further generalized with a proper gluon propagator and starting from QCD in \cite{fra0}. On a Nambu-Jona-Lasinio model bosonizationc can be applied to introduce meson fields from quark fields \cite{ebe}. Bosonization techniques make comparable Nambu-Jona-Lasinio models and ChPT \cite{kle2}.

Not less relevant has been the approach of quantum spectral sum rules that have taken the start from the pioneer work of Vanshtein, Shifman and Zakharov \cite{vsz1,vsz2} that also pointed out the existence of gluon and quark condensates. Sum rule techniques showed the need for a further meson with a mass around a GeV \cite{nar1}.
Bosonization procedures produce the appearance of the field of the $\sigma$ meson in the Nambu-Jona-Lasinio model and this field is needed in ChPT. Indeed, existence of this particle has been the argument of longstanding discussion and, after an initial insertion, it was dismissed by Particle Data Group for over twenty years. Problem in experiments arisen from the fact that this resonance is quite broad and so difficult to be properly measured. Today there is widely acceptance and a decisive step forward is due to the work of Caprini, Colangelo and Leutwyler \cite{lcc} further supported by the work of Yndurain, Garcia-Martin and Pelaez \cite{pel}. The main question now moved from the existence of this resonance to its nature. In this paper we show evidence toward a large gluonic component in it.

Notwithstanding a wide success encountered by these techniques, the understanding of the lower part of the spectrum of QCD escaped us and is reason of a very hot debate in these days \cite{cre}. The reason has been already pointed above and relies on our difficulties to manage QCD at very low energies. Lattice computations have given very few help about and it is possible that there is something deep to be better understood with this approach.

So far, a fundamental question remains unanswered. There has been no approach to derive a low-energy limit directly from QCD. There has been no successful and widely accepted proposal so far. But a significant track has been unveiled by Goldman and Haymaker \cite{tg}. These authors proved that QCD has a meaningful low-energy limit provided the form of the gluon propagator is given and is not as singular as the propagator of a free massless particle in the limit of lower momenta. On a same track, one can do a current expansion into the generating functional of QCD recovering Goldman and Haymaker setback \cite{rob}. At some point one should know the gluon propagator to recover a low-energy limit from QCD.

The question of the understanding of the gluon propagator in the low-energy limit has been matter of wide debate in the last few years, for a pure Yang-Mills theory. There has been groundbreaking theoretical work, well recounted in \cite{avs}, claiming that, lowering momenta, the gluon propagator should have gone to zero. Similarly, as a consequence, the running coupling should have reached a fixed point. People working on the lattice, at first, were not able to have a clear understanding about the right behavior and, for some years, appeared as this theoretical scenario was supported. A crucial step happened at the XXV International Symposium on Lattice Field Theory, held at Regensburg in Germany on 2007, where two groups announced results on very large lattices \cite{cuc,ste}. Results on lattice showed unequivocally that the gluon propagator, at very low momenta, reaches a finite value different from zero, with a curve resting on a plateau for the Landau gauge. Running coupling is seen to tend toward zero  having no fixed point at all. This latter result is in agreement with the beta function for a SUSY Yang-Mills theory that is exactly known \cite{vsz3} and its extrapolation to Yang-Mills theory \cite{san,boc}.

In this paper we will derive the low-energy limit of QCD, based on a sound derivation of the gluon propagator in the infrared. Several applications to low-energy QCD processes will be given. The paper is so structured. In sec.\ref{sec3} we introduce our notations for QCD. In sec.\ref{sec4} we perform the analysis of the low-energy limit of QCD obtaining the gluon propagator and showing how to overcome Goldman and Haymaker setback. In sec.\ref{sec5} we gives some applications to low-energy processes in QCD where glue excitations and quarks interact at tree level. Finally, in sec.\ref{sec6} conclusions are presented.

\section{\label{sec3} Quantum Chromodynamics}

In order to fix notation, we give below the formulation of QCD we will work with.

The action for QCD can be written as
\begin{equation}
   S_{YM} = -\int d^4x\left[\frac{1}{4}G^a_{\mu\nu}G^{a\mu\nu}
   +\frac{1}{2\alpha}(\partial\cdot A)^2\right]
\end{equation}
being
\begin{equation}
   G^a_{\mu\nu}=\partial_\mu A^a_\nu-\partial_\nu A^a_\mu+gf^{abc}A^b_\mu A^c_\nu
\end{equation}
the field tensor and $g$ the coupling constant that is dimensionless in four dimensions. $f^{abc}$ are the structure constants of SU(3) group with the generators $[\lambda_a,\lambda_b]=if^{abc}\lambda_c$. Here $a,b,c=1,2,3$. Finally, the value of $\alpha$ gives the fixing gauge term. 

For quarks one has
\begin{equation}
   S_m=\int d^4x\sum_q\bar q\left(i\gamma\cdot\partial 
   + g\frac{\lambda^a}{2}\gamma\cdot A^a - m_q\right)q.
\end{equation}
being $q=u,d,s,\ldots$ the quark fields and $m_q$ the corresponding masses. 

Finally, we have to introduce Fadeev-Popov ghost. This is done in the following way
\begin{equation}
    S_g = -\int d^4x\left[\partial^\mu\bar c^a\partial_\mu c^a
    +gf^{abc}\partial_\mu\bar c^a A^{b\mu}c^c\right].
\end{equation}

From these equations we can formulate the generating functional for quantum field theory as
\begin{eqnarray}
\label{eq:gf}
   Z[\eta,\bar\eta,\epsilon,\bar\epsilon,j]=\int \prod_q[dq][d\bar q][dc][d\bar c][dA]
   e^{i(S_{YM}+S_m+S_g)} \times&& \\ \nonumber
   e^{i\int d^4x\sum_q[\bar q(x)\eta_q(x)+\bar\eta_q(x)q(x)]} 
   e^{i\int d^4x[\bar c^a(x)\epsilon^a(x)+\bar\epsilon^a(x)c^a(x)]} 
   e^{i\int d^4x j^a_\mu(x)A^{a\mu}(x)}.&&
\end{eqnarray}
here and in the following $\hbar=c=1$. Our aim in this paper will be to get a proper characterization of the behavior of the quantum theory in the low-energy limit. 

\section{\label{sec4} Analysis of low-energy limit}

\subsection{General considerations}

Quantum field theory relies heavily on our ability to find out exact solutions to classical equations of motion. Once such solutions are known, we are able to do perturbation theory and obtaining results to be compared with experiments. It is interesting to note that the kind of perturbation series will depend on the nature of the exact solutions we will work with. It is also possible to work with approximate solutions as starting point for perturbation theory. In this case one must be sure to work consistently and be able to get systematically all higher order terms.

A possibility to formulate a strong coupling expansion in quantum field theory was put forward in \cite{fra1}. In this case we were able to apply this approach to a quartic scalar field theory obtaining the corresponding propagator and the mass spectrum. The theory is seen to be trivial as, at lower momenta, the coupling goes to zero \cite{fra11,sus1,sus2}. The corresponding expansion is indeed a strong coupling expansion as the parameter of the series is the inverse of the coupling.

Due to the peculiar nature of a quartic field theory and a simple observation that Yang-Mills theory seems a vectorial generalization of this theory, one may think that a relation should exist between these twos. Indeed, there exists a class of exact solutions that are common to the classical equations of these theories but these solutions can only depend on time variable. When a gradient expansion is used to correct these solutions one observe that this mapping is only perturbative in the inverse of the couplings. Couplings are connected by the relation $\lambda=Ng^2$ being $\lambda$ the coupling of the scalar theory and $g$ the coupling of the SU(N) Yang-Mills theory. This is the content of the so-called mapping theorem that we were able to prove in \cite{fra12,fra13}. The relevance of this result relies on the fact that we can pass all the results of the scalar field theory to the Yang-Mills theory when a strong coupling limit is considered. In this way we are able to get both the gluon propagator and the mass spectrum in the same limit and Yang-Mills theory is seen to display a mass gap.

An immediate by-product of the mapping theorem is the opportunity, in the vein of Goldman and Haymaker \cite{tg}, to obtain the correct low-energy limit of QCD with all the parameters properly fixed by the theory. This limit is a Nambu-Jona-Lasinio model equivalent to a Yukawa model in a proper approximation. In the next sections we will exploit all this in a consistent mathematical way. Our main goal will be to put on a sound basis the techniques to analyze in a closed form low-energy QCD. 

\subsection{Exact solutions and mapping theorem}

Quite recently, we showed that a gradient expansion is indeed a strong coupling expansion\cite{fra14}. This can be seen very easily for a quartic scalar field theory. One has
\begin{equation}
   \Box\phi+\lambda\phi^3=0
\end{equation}
with $\lambda\rightarrow\infty$. Let us rescale the time variable as $\tau=\sqrt{\lambda}t$ and take
\begin{equation}
   \phi=\phi_0+\frac{1}{\lambda}\phi_1+O\left(\frac{1}{\lambda^2}\right).
\end{equation}
We get a consistent set of differential equations
\begin{eqnarray}
   \partial_t^2\phi_0+\phi_0^3&=&0 \\
   \partial_t^2\phi_1+3\phi_0^2\phi_1&=&\Delta_2\phi_0 \\
   &\ldots&
\end{eqnarray}
that is clearly a gradient expansion. This means that gradient expansion is nothing else than a strong coupling expansion \cite{fra14}. Being this a gradient expansion, it is not difficult to note that an exact solution to our equation of motion is obtained if only time dependence is retained solving just the first equation of the given set. Moreover, this will produce an exact solution to the full equation after a Lorentz boost. Such a solution can be written down as
\begin{equation}
\label{eq:td}
   \phi(t,0) = \mu\left(\frac{2}{\lambda}\right)^\frac{1}{4}
   {\rm sn}\left(\left(\frac{\lambda}{2}\right)^\frac{1}{4}\mu t+\theta,i\right)
\end{equation}
being $\mu$ and $\theta$ integration constants. A Lorentz boost takes this solution to
\begin{equation}
\label{eq:td1}
   \phi(x) = \mu\left(\frac{2}{\lambda}\right)^\frac{1}{4}{\rm sn}(p\cdot x+\theta,i)
\end{equation}
provided $p^2=\left(\frac{\lambda}{2}\right)^\frac{1}{2}\mu^2$. Now, given Yang-Mills equations
\begin{equation}
\label{eq:ym}    
    \partial^\mu\partial_\mu A^a_\nu-\left(1-\frac{1}{\alpha}\right)
    \partial_\nu(\partial^\mu A^a_\mu)
    +gf^{abc}A^{b\mu}(\partial_\mu A^c_\nu-\partial_\nu A^c_\mu)
    +gf^{abc}\partial^\mu(A^b_\mu A^c_\nu)
    +g^2f^{abc}f^{cde}A^{b\mu}A^d_\mu A^e_\nu = 0
\end{equation}
it is always possible to choice a set of constants $\eta^a_\mu$, Smilga's choice \cite{smi}, such that the solution (\ref{eq:td}) is also a solution of these equations. But here we cannot just boost these solutions to get something like (\ref{eq:td1}). Rather, due to the presence of an arbitrary choice of gauge, we can only state the following perturbative mapping between the scalar field and Yang-Mills potentials
\begin{equation}
    A_\mu^a(x)=\eta^a_\mu\phi(x)+O\left(\frac{1}{\sqrt{N}g}\right)
\end{equation}
provided $\lambda=Ng^2$. This is exactly the content of the {\sl mapping theorem} \cite{fra12,fra13}. Due to the nature of these approximate solutions of the Yang-Mills equations, they indeed provide the behavior of the classical theory when the coupling becomes increasingly large. This is exactly the limit we are interested in. So, a quantum field theory can be built on these solutions through a perturbation theory in the strong coupling limit.

\subsection{Gluon propagator}

In order to work out a low-energy limit for QCD we will apply the mapping theorem to the generating functional (\ref{eq:gf}). This will give
\begin{eqnarray}
\label{eq:yu}
Z[\eta,\bar\eta,j]=\int \prod_q[dq][d\bar q][d\phi]
e^{i(N^2-1)\int d^4x\left[\frac{1}{2}(\partial\phi)^2-\frac{Ng^2}{4}\phi^4\right]}\times&& \\ \nonumber
e^{i\int d^4x\sum_q\bar q(x)\left[\gamma\cdot\left(i\partial+g\frac{\lambda\cdot\eta}{2}\phi\right)-m_q\right]q(x)}
e^{i\int d^4x\sum_q[\bar q(x)\eta_q(x)+\bar\eta_q(x)q(x)]}\times && \\ \nonumber 
e^{i\int d^4x j_\phi(x)\phi(x)}+O(1/\sqrt{N}g)&&
\end{eqnarray}
having set $j_\phi=\eta\cdot j$.
It is interesting to note that, at this order of the strong coupling expansion, the ghost field decouples and can be ignored in our analysis, this is in agreement with lattice results\cite{cuc,ste}. Higher order terms can be properly computed through a gradient expansion of the classical equations of motion. We can recognize here the quantum field theory of a Yukawa model. This model can be cast into a Gaussian form and so can be reduced to a manageable form. So, let us consider the generating functional for the scalar field
\begin{equation}
   Z_0[j]=\int d[\phi] e^{i\int d^4x
   \left[\frac{1}{2}(\partial\phi)^2-\frac{\lambda}{4}\phi^4+j\phi\right]}
\end{equation}
and take the limit $\lambda\rightarrow\infty$. For a strong coupling expansion, we separate gradients from other terms as
\begin{equation}
   Z_0[j]=\int d[\phi] e^{i\int d^4x
   \left[\frac{1}{2}(\partial_t\phi)^2-\frac{\lambda}{4}\phi^4+j\phi\right]}
   e^{-i\int d^4x\left[\frac{1}{2}(\nabla\phi)^2\right]}
\end{equation}
and finally, we have to solve
\begin{equation}
   \partial_t^2\phi_c+\lambda\phi_c^3=j.
\end{equation}
Taking the expansion
\begin{equation}
   \phi=\phi_c+\delta\phi
\end{equation}
the generating functional is reduced to the form \cite{fra1}
\begin{equation}
   Z_0[j]=e^{\frac{i}{2}\int d^4xj\phi_c}F[j]
\end{equation}
with the functional $F[j]$ containing terms negligible in the limit $\lambda\rightarrow\infty$. Another approximation to be applied is to take, in a small time limit \cite{fra2,fra3},
\begin{equation}
   \phi_c(x)\approx\int d^4x'G(x-x')j(x'),
\end{equation}
also to be seen as a current expansion\cite{rob}, being
\begin{equation}
   G(x-x')=\delta^3(x-x')\tilde G(t-t')
\end{equation}
and
\begin{equation}
   \partial_t^2\tilde G(t-t')+\lambda\tilde G(t-t')^3=\delta(t-t')
\end{equation}
that can be exactly solved giving
\begin{equation}
\label{eq:G}
   \tilde G(t-t')=\theta(t-t')\mu\left(\frac{2}{\lambda}\right)^\frac{1}{4}{\rm sn}
   \left(\left(\frac{\lambda}{2}\right)^\frac{1}{4}\mu t,i\right)
\end{equation}
and the corresponding time-reversed solution. This gives finally a Gaussian generating functional that holds for a strongly coupled theory, the infrared limit,
\begin{equation}
   Z_0[j]\approx e^{\frac{i}{2}\int d^4xd^4yj(x)\Delta(x-y)j(y)}
\end{equation}
and the propagator given by
\begin{equation}
   \Delta(x-y)=\delta^3(x-y)\left[\tilde G_+(t_x-t_y)+\tilde G_-(t_x-t_y)\right]
\end{equation}
with $G_+$ given in eq.(\ref{eq:G}) and $G_-$ the corresponding time reversed, $t\rightarrow -t$, Green function. Now, we are able to obtain the complete propagator of this quantum field theory in the infrared limit. The relevance of this result relies on the fact that, due to the mapping theorem, this will also represent the infrared gluon propagator for Yang-Mills theory. Indeed, we note that for the Jacobi snoidal elliptical function the following identity holds\cite{gr}
\begin{equation}
    {\rm sn}(u,i)
    =\frac{2\pi}{K(i)}\sum_{n=0}^\infty\frac{(-1)^ne^{-(n+\frac{1}{2})\pi}}{1+e^{-(2n+1)\pi}}
    \sin\left[(2n+1)\frac{\pi u}{2K(i)}\right]
\end{equation}
being $K(i)$ the constant
\begin{equation}
    K(i)=\int_0^{\frac{\pi}{2}}\frac{d\theta}{\sqrt{1+\sin^2\theta}}\approx 1.3111028777.
\end{equation}
So, one has
\begin{equation}
\label{eq:prop}
    \Delta(\omega,0)=\sum_{n=0}^\infty\frac{B_n}{\omega^2-m_n^2+i\epsilon}
\end{equation}
being
\begin{equation}
    B_n=(2n+1)\frac{\pi^2}{K^2(i)}\frac{(-1)^{n+1}e^{-(n+\frac{1}{2})\pi}}{1+e^{-(2n+1)\pi}},
\end{equation}
and,
\begin{equation}
\label{eq:ms}
    m_n = \left(n+\frac{1}{2}\right)\frac{\pi}{K(i)}
    \left(\frac{\lambda}{2}\right)^{\frac{1}{4}}\mu
\end{equation}
the spectrum of the theory. This theory displays a mass gap in the infrared limit and this also must happen to Yang-Mills theory. Now, we can boost this propagator from the rest frame and this will amount to sum all the gradient terms providing in the end
\begin{equation}
\label{eq:prop1}
    \Delta(p)=\sum_{n=0}^\infty\frac{B_n}{p^2-m_n^2+i\epsilon}.
\end{equation}
So, we are in a position to write down the infrared propagator for SU(N) Yang-Mills theory in the Landau gauge as
\begin{equation}
   D_{\mu\nu}^{ab}(x-y)=\delta_{ab}\left(\eta_{\mu\nu}-\frac{p_\mu p_\nu}{p^2}\right)\Delta(p)
\end{equation}
provided we take $\lambda=Ng^2$. This is a key result, as it overcomes Goldman and Haymaker setback, and we are able to get a meaningful low energy limit for QCD.

\subsection{Low-energy QCD}

Once we know the gluon propagator, we can do a change of coordinates into the generating functional of QCD. This change of coordinates is the one suggested in the analysis of the scalar theory. So, in the Yukawa model (\ref{eq:yu}) we put
\begin{equation}
   \phi(x)\approx\int d^4y\Delta(x-y)\left[j_\phi(y)
   +g\sum_q\bar q(y)\frac{\gamma\lambda\cdot\eta}{2}q(y)\right],
\end{equation}
and we use the reduction to a Gaussian form exploited in the preceding section. This will give the generating functional
\begin{eqnarray}
\label{eq:yu1}
Z[\eta,\bar\eta,j]=e^{i(N^2-1)\int d^4xd^4y\left[\frac{1}{2}j_\phi(x)\Delta(x-y)j_\phi(y)\right]}\int \prod_q[dq][d\bar q]
\times&& \\ \nonumber
e^{i\int d^4x\sum_q\bar q(x)
\left[\gamma\cdot\left(i\partial+g\frac{\lambda\cdot\eta}{2}
\int d^4y\Delta(x-y)\left[j_\phi(y)+\frac{1}{2}g\sum_q\bar q(y)\frac{\gamma\lambda\cdot\eta}{2}q(y)\right]\right)-m_q\right]q(x)}
e^{i\int d^4x\sum_q[\bar q(x)\eta_q(x)+\bar\eta_q(x)q(x)]}\times && \\ \nonumber 
+O(1/\sqrt{N}g).&&
\end{eqnarray}
This result is quite striking. The Yukawa model we obtained through the mapping theorem just gives rise to a modified Nambu-Jona-Lasinio model, without contact interaction, plus a term describing the pure interaction of the quark fields with a set of scalar excitations. These excitations should represent the possible observed glueball spectrum. For completeness, we give below the Lagrangian of the model, neglecting the Gaussian term describing free scalar excitations,
\begin{eqnarray}
   L&=& \sum_q\bar q(x)\left(i\gamma\cdot\partial-m_q\right)q(x) \\ \nonumber
    &+&\frac{g}{2}\sum_q\bar q(x)\lambda^a\eta^a_\mu\gamma^\mu q(x)
    \int d^4y\Delta(x-y)j_\phi(y) \\ \nonumber
    &+&\frac{1}{2}\frac{g^2}{4}\sum_{q,q'}\bar q(x)\lambda^a\eta^a_\mu\gamma^\mu q(x)
      \int d^4y\Delta(x-y)\bar q'(y)\lambda^b\eta^b_\nu\gamma^\nu q'(y) \\ \nonumber
    &+&O(1/\sqrt{N}g)
\end{eqnarray}
This Lagrangian, together with the given generating functional, fixes univocally the behavior of QCD at lower energies. It is interesting to note that we have no yet done a Nambu-Jona-Lasinio point contact approximation. So, we can be sure that the theory is renormalizable. The form of the gluon propagator grants that a point contact approximation can always be done recovering a fully Nambu-Jona-Lasinio model removing Goldman and Haymaker setback. Confinement is granted again by the gluon propagator that displays a full set of massive excitations. When we do a reduction to the Nambu-Jona-Lasinio model with a point contact interaction we lose both renormalizability and confinement but breaking of the chiral symmetry is maintained.

\subsection{Nambu-Jona-Lasinio limit}

Nambu-Jona-Lasinio limit corresponds to current-current approximation in the low-energy field theory we obtained so far. Existence of this limit can be easily inferred from the gluon propagator (\ref{eq:prop1}) that has a finite limit when momenta go to zero. One has
\begin{equation}
   \Delta(x-y)\approx \frac{3.76}{\sigma}\delta^4(x-y)
\end{equation}
being $\sigma$ the string tension as defined above. From lattice computations one has $\sqrt{\sigma}=0.41\div 0.44\ GeV$ but this value should be obtained from experiment. Now, we define
\begin{equation}
   G_{NJL}=\frac{3.76}{\sigma}g^2=3.76\frac{4\pi\alpha_s}{\sigma}
\end{equation}
and
\begin{equation}
   G'_{NJL}=\frac{3.76}{\sigma}g=3.76\frac{\sqrt{4\pi\alpha_s}}{\sigma}.
\end{equation}
The corresponding Lagrangian will be
\begin{eqnarray}
   L_{NJL}&=& \sum_q\bar q(x)\left(i\gamma\cdot\partial-m_q\right)q(x) \\ \nonumber
    &+&\frac{1}{2}G'_{NJL}\sum_q\bar q(x)
    \lambda^a\eta^a_\mu\gamma^\mu q(x)j_\phi(x) \\ \nonumber
    &+&\frac{1}{2}\frac{G_{NJL}}{4}\sum_{q,q'}\bar q(x)\lambda^a\eta^a_\mu\gamma^\mu q(x)
      \bar q'(x)\lambda^b\eta^b_\nu\gamma^\nu q'(x) \\ \nonumber
    &+&O(1/\sqrt{N}g)
\end{eqnarray}
producing a generating functional
\begin{eqnarray}
\label{eq:yu2}
Z_{NJL}[\eta,\bar\eta,j]&=&e^{i(N^2-1)\int d^4xd^4y\left[\frac{1}{2}j_\phi(x)\Delta(x-y)j_\phi(y)\right]}
\int\prod_q[dq][d\bar q]\times \\ \nonumber
&& e^{i\int d^4x\left[\sum_q\bar q(x)\left(i\gamma\cdot\partial-m_q\right)q(x)
    +\frac{1}{2}G'_{NJL}\sum_q\bar q(x)\lambda^a\eta^a_\mu\gamma^\mu q(x)j_\phi(x)
    +\frac{1}{2}\frac{G_{NJL}}{4}\sum_{q,q'}\bar q(x)\lambda^a\eta^a_\mu\gamma^\mu q(x)
     \bar q'(x)\lambda^b\eta^b_\nu\gamma^\nu q'(x)\right]}\times \\ \nonumber
&& e^{i\int d^4x\sum_q[\bar q(x)\eta_q(x)+\bar\eta_q(x)q(x)]}+O(1/\sqrt{N}g).
\end{eqnarray}


At this stage we take a current expansion\cite{rob} as $j_q j_\phi$, $j_q^2$ and so on and we are left with a first order approximation in the strong coupling limit
\begin{eqnarray}
\label{eq:yu3}
Z_0[\eta,\bar\eta,j]&\approx&e^{i(N^2-1)\int d^4xd^4y\left[\frac{1}{2}j_\phi(x)
\Delta(x-y)j_\phi(y)\right]}\int\prod_q[dq][d\bar q]
\times \\ \nonumber
&& e^{i\int d^4x\left[\sum_q\bar q(x)\left(i\gamma\cdot\partial-m_q\right)q(x)
    +\frac{1}{2}G'_{NJL}\sum_q\bar q(x)\lambda^a\eta^a_\mu\gamma^\mu q(x)j_\phi(x)
    \right]}\times \\ \nonumber
&& e^{i\int d^4x\sum_q[\bar q(x)\eta_q(x)+\bar\eta_q(x)q(x)]}.
\end{eqnarray}
where only the term $j_q j_\phi$ is retained. This approximation amounts to ignore quark loops and to consider only quark-gluon vertexes. We can perform the integral if we know the solution to the equation
\begin{equation}
   \left[i\gamma\cdot\partial-m_q+\frac{1}{2}G'_{NJL}
   \lambda^a\eta^a_\mu\gamma^\mu j_\phi(x)\right]S_q[j_\phi;x-x']=\delta^4(x-x').
\end{equation}
The idea at this stage is to perform again a gradient expansion retaining just the leading term. One gets without difficulty the following result
\begin{equation}
    S[j_\phi,x-x']=\theta(t_x-t_{x'})\delta^3(x-x')\exp\left[iG'_{NJL}\frac{\lambda^a}{2}
    \gamma_0\gamma_i\eta_i^a\int_{t_x}^{t_{x'}} dt'j_\phi(t'-t_{x'},x-x')\right].
\end{equation}
Finally, we have the low-energy generating functional for QCD as
\begin{equation}
\label{eq:leqcd}
    Z_0[\eta,\bar\eta,j_\phi]\approx
    \exp\left\{\frac{i}{2}(N^2-1)\int d^4xd^4y j_\phi(x)\Delta(x-y)j_\phi(y)\right\}
    \exp\left\{i\int d^4xd^4y\sum_q\bar\eta_q(x)S[j_\phi,x-y]\eta_q(y)\right\}.
\end{equation}
This generating functional describes the behavior of glue scalar excitations and their interaction with quarks. To account for loop contributions due to quarks and scalar excitations one has to consider again quark current-current interaction.

\section{\label{sec5} QCD processes at lower energies}

\subsection{Narison-Veneziano sum rules}

In the eighties, Narison and Veneziano derived some sum rules for scalar resonances in QCD \cite{nv1}. These rules are relevant to obtain some properties, as the decay widths, of such particles and give hints for their observation. In the following we prove that Narison-Veneziano sum rules are properly verified for the spectrum of $0^{++}$ mesons decaying in $\pi^+\pi^-$ obtaining in this way some relevant information about this decay but, more importantly, we get the form factors of scalar resonances. 

In order to obtain this result we use the leading order approximation of an expansion in $1/\sqrt{3}g$ for QCD. Firstly, we consider
\begin{equation}
\label{eq:nv2}
   \frac{1}{4}\sum_Sg_{S\pi^+\pi^-}\frac{\sqrt{2}f_S}{M^2_s}=1
\end{equation}
and from this we get the corresponding value of $\alpha_s$. We use this to finally check that
\begin{equation}
\label{eq:nv1}
    \frac{1}{4}\sum_Sg_{S\pi^+\pi^-}\sqrt{2}f_S\approx 0.
\end{equation}
In order to prove this, we have to compute $f_S$ that are the decay form factors for scalar mesons. This is easily accomplished through the mapping theorem that gives
\begin{equation}
   A_\mu^a(x)=\eta_\mu^a\phi(x)+O(1/\sqrt{3}g)
\end{equation}
being
\begin{equation}
   \Box\phi+3g^2\phi=0.
\end{equation}
We recognize here the one-point function of the quantum field theory when all the currents are set to zero. This equation has the following exact solution
\begin{equation}
   \phi(x)=\sqrt{\frac{\sigma}{6\pi\alpha_s}}{\rm sn}(p\cdot x+\theta,i)
\end{equation}
being sn a Jacobi elliptic function and provided that
\begin{equation}
   p^2=\sigma
\end{equation}
being $\sqrt{\sigma}=\Lambda(3g^2/2)^\frac{1}{4}$ the string tension taken to be 410 to 440 MeV in lattice computations. But this solution can be also given as the sum of free massive excitations through its Fourier series as
\begin{equation}
   \phi(t,0)=\sqrt{\frac{\sigma}{6\pi\alpha_s}}\frac{2\pi}{K(i)}
   \sum_{n=0}^{\infty}(-1)^n
   \frac{e^{-\left(n+\frac{1}{2}\right)\pi}}{1+e^{-(2n+1)\pi}}e^{-im_nt+\theta}+c.c.
\end{equation}
with the meson spectrum (see (\ref{eq:ms}))
\begin{equation}
   M_{S_n}=(2n+1)\frac{\pi}{2K(i)}\sqrt{\sigma}
\end{equation}
and the form factors
\begin{equation}
\label{eq:fsn}
   f_{S_n}=\sqrt{\frac{\sigma}{6\pi\alpha_s}}\frac{2\pi}{K(i)}(-1)^n
   \frac{e^{-\left(n+\frac{1}{2}\right)\pi}}{1+e^{-(2n+1)\pi}}
\end{equation}
where we see that they depend on $1/\sqrt{\alpha_s}$ that is clearly not obtainable from a standard small perturbation approach. It is interesting to note that they have alternating signs in agreement with what one should expect from Narison-Veneziano sum rules.

Finally, putting all together and doing the severe approximation that all $g_{S\pi^+\pi^-}$ are equal, evaluating the sum will give from eq.(\ref{eq:nv2})
\begin{equation} 
   g_{S\pi^+\pi^-}\approx (6\pi\alpha_s)^\frac{1}{2}\frac{5\pi}{2\sqrt{2}K(i)}\sqrt{\sigma}
\end{equation}
that, taking $g_{S\pi^+\pi^-}\approx 3.8\div 3.9 GeV$, will give
\begin{equation}
   \alpha_s=0.25\div 0.26 
\end{equation}
a fairly good value. To complete our check, we consider the other Narison-Veneziano sum rule to obtain
\begin{equation}
   g_{S\pi^+\pi^-}\frac{\sqrt{\sigma}}{(6\pi\alpha_s)^\frac{1}{2}}0.024\approx 0.02\ GeV^2
\end{equation}
and so we see that both sum rules are well verified for $0^{++}$ glue excitations.

\subsection{Properties of the $\sigma$ meson and higher scalar excitations}

Several processes involving glue scalar excitations can be computed through the generating functional (\ref{eq:leqcd}). The mapping theorem showed that the glue sector of the QCD action produces a set of scalar excitations with a mass spectrum
\begin{equation}
       m_n = (2n+1)\frac{\pi}{2K(i)}\sqrt{\sigma}
\end{equation}
being
\begin{equation}
       \sigma=\sqrt{6\pi\alpha_s}\mu^2
\end{equation}
the string tension. From these considerations, we observe that the spectrum displays a massive state with a mass around 500 MeV. Such a state has been observed and some phenomenological analysis strongly support its existence \cite{lcc,pel}. This state is identified as the $\sigma$ meson. This resonance is broad and has a prevailing decaying mode in two pions and a very small decay rate in two photons. This latter decay has been properly understood as a rescattering effect \cite{nar2}. In the debate about the nature of this resonance there have been some works pointing out its mostly glue content \cite{nar2,nar3,nar4}. We will give a significant support to these results.

\begin{figure}[tbp]
\begin{center}
\includegraphics[width=320pt]{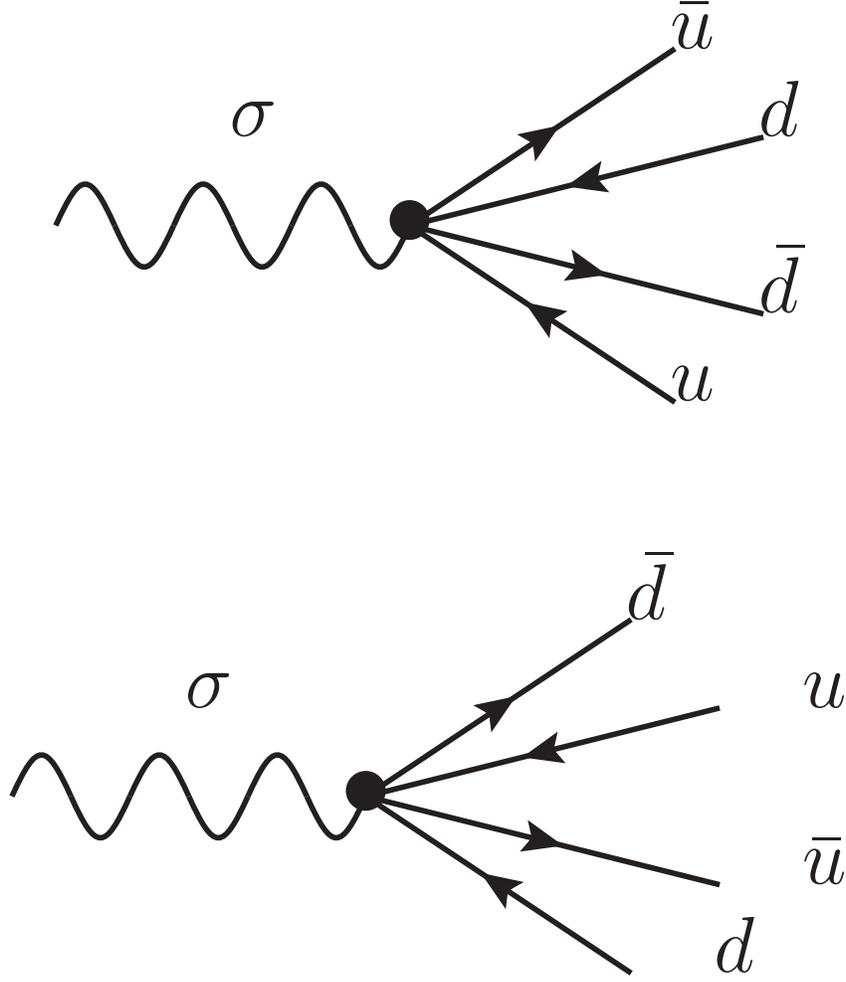}
\caption{\label{fig:fig1} $\sigma\rightarrow\pi^++\pi^-$ amplitudes.}
\end{center}
\end{figure}

The process
\begin{equation}
   \sigma\rightarrow\pi^++\pi^-
\end{equation}
can be described by the Feynman diagrams given in fig.\ref{fig:fig1}.
This is a vertex process and the amplitudes are given by
\begin{eqnarray}
    &&\left.i\frac{\delta}
    {\delta j_\phi(x)}\frac{\delta}{i\delta\bar\eta_u(x)}\frac{i\delta}{\delta\eta_d(x)}
    \frac{\delta}{i\delta\bar\eta_d(x)}
    \frac{i\delta}{\delta\eta_u(x)}Z[\eta,\bar\eta,j_\phi]\right|_{j_\phi,\bar\eta,\eta=0}
    = \\ \nonumber
    &-&iG'_{NJL}\frac{\lambda^a}{2}\gamma_0\gamma_i\eta_i^a \times \\ \nonumber
    &&\left[\theta(t_4-t_3)\theta(t_2-t_5)\theta(t_2-t_5-t_1)\delta^3(x_2-x_5)
    \delta^3(x_4-x_3)\delta^3(x_1)\right. \\ \nonumber
    &+&\left.\theta(t_4-t_3)\theta(t_2-t_5)\theta(t_4-t_3-t_1)
    \delta^3(x_2-x_5)\delta^3(x_4-x_3)\delta^3(x_1)\right].
\end{eqnarray}
In order to evaluate this vertex we use the following expression for the Heaviside function
\begin{equation}
   \theta(t)=-\frac{1}{2\pi i}\int_{-\infty}^{+\infty}dE\frac{1}{E+i0}e^{-iEt}.
\end{equation}
We have to go from this n-point function to the probability amplitude introducing pion fields. We do this with the following rule: we remove the contributions from the Heaviside functions with the product $m_\sigma f_\pi^2$, assuming $\sigma$ at rest and $m_\sigma$ its mass and $f_\pi$ the pion decay constant taken to be 130 MeV. This corresponds to LSZ reduction. So, when we take the trace of the square of the amplitude, we will be left with
\begin{equation}
   |{\cal M}_{if}|^2=4(N^2-1){G'}^{2}_{NJL} m_\sigma^2f_\pi^4
\end{equation}
that we specialize to $N=3$. This will give the rate
\begin{equation}
   \Gamma_\sigma = \frac{2}{\pi}{G'}^{2}_{NJL} 
   m_\sigma f_\pi^4\sqrt{1-\frac{4m_\pi^2}{m_\sigma^2}}
\end{equation}
that is in agreement with recent derivations from experiments \cite{pel} being $\Gamma_\sigma/2=255\pm 10\ MeV$ for a mass $m_\sigma=484\pm 17\ MeV$ when $\alpha_s\approx 0.59$. A similar conclusion can be drawn also with respect to the derivation given in \cite{lcc}, presenting a mass $m_\sigma=441^{+16}_{-8}\ MeV$ and $\Gamma_\sigma/2=279^{+9}_{-12.5}\ MeV$, for $\alpha_s\approx 0.47$ showing that both results are consistent each other with respect to physical values of the strong coupling constant. In these computations we have fixed the string tension through the mass formula for the $\sigma$ resonance obtained putting $n=0$ into eq.(\ref{eq:ms}).

From this result, we are able to evaluate the coupling as given in literature \cite{nar2,nar3}. One sets \cite{nar3}
\begin{equation}
\Gamma_\sigma=\frac{|g_{\sigma\pi\pi}|^2}{16\pi m_\sigma}\sqrt{1-\frac{4m_\pi^2}{m_\sigma^2}}
\end{equation}
being $g_{\sigma\pi\pi}$ the coupling. This gives
\begin{equation}
     |g_{\sigma\pi\pi}|=4\sqrt{2}G'_{NJL} m_\sigma f_\pi^2\approx 1.68\div 1.93\ GeV
\end{equation}
when the parameter given in \cite{pel,lcc} are used. As one should expect, this number is quite sensible to the value of the $\sigma$ mass and the running coupling.

In order to have a comparison with the observed spectrum of Yang-Mills theory on the lattice \cite{tep,mor}, we consider the standard two-point function
\begin{equation}
   C^{ab}_{\mu\nu}(t-t')=\langle A_\mu^a(t,0)A_\nu^b(t',0)\rangle
\end{equation}
having used translational invariance. Now, we apply the mapping theorem and we have in the end
\begin{equation}
   C^{ab}_{\mu\nu}(t-t')=\eta_\mu^a\eta_\nu^b\langle\phi(t,0)\phi(t',0)\rangle+O(1/\sqrt{N}g).
\end{equation}
This result is rather beautiful as we can use the exact classical solutions of the scalar field. Indeed, this function solves the equation $\partial_t^2\langle\phi(t,0)\phi(t',0)\rangle+\lambda\langle\phi(t,0)\phi(t',0)\rangle^3=\delta(t-t')$ and so we can conclude that
\begin{equation}
   C^{ab}_{\mu\nu}(t-t')=\theta(t-t')\eta_\mu^a\eta_\nu^b\sum_n(-1)^n
   \frac{\pi}{iK(i)}\frac{e^{-\left(n+\frac{1}{2}\right)\pi}}
   {1+e^{-(2n+1)\pi}}e^{-im_n(t-t')}+c.c.+O(1/\sqrt{N}g)
\end{equation}
as expected from lattice computations where the spectrum is given by 
\begin{equation}
\label{eq:ms1}
    m_n = \left(n+\frac{1}{2}\right)\frac{\pi}{K(i)}\sqrt{\sigma}
\end{equation}
being $\sqrt{\sigma}=\left(\frac{Ng^2}{2}\right)^{\frac{1}{4}}\mu$ the root of string tension. So, the mapping theorem grants an identical excitation spectrum for a pure Yang-Mills theory and a quartic massless scalar field theory in the infrared. This has been confirmed quite recently on lattice computations for d=2+1 \cite{raf}. In d=3+1, for comparison with \cite{tep}, we get the following table
\begin{center}
\begin{table}[h]
\begin{tabular}{|c|c|c|c|} \hline\hline
Excitation & Lattice & Theoretical & Error \\ \hline
$\sigma$   & -       & 1.198140235 & - \\ \hline 
0$^{++}$   & 3.55(7) & 3.594420705 & 1\% \\ \hline
0$^{++*}$  & 5.69(10)& 5.990701175 & 5\% \\ \hline
$\sigma^*$ & -       & 2.396280470 & - \\ \hline 
2$^{++}$   & 4.78(9) & 4.792560940 & 0.2\% \\ \hline
2$^{++*}$  & -       & 7.188841410 & - \\ \hline\hline
\end{tabular}
\end{table}
\end{center}
where our numbers are computed with the formulas
\begin{eqnarray}
   \frac{m_n}{\sqrt{\sigma}} &=& \left(n+\frac{1}{2}\right)\frac{\pi}{K(i)} \\ \nonumber
   \frac{m_{n,m}^*}{\sqrt{\sigma}} &=& \left(n+m+1\right)\frac{\pi}{K(i)} \\ \nonumber
   &\vdots&
\end{eqnarray}
that are pure numbers (see also \cite{cmn}). 

Finally, we note that a decay $S_n\rightarrow\pi^++\pi^-$ is a clear signature for these scalar excitations. This should explain why it is so difficult to identify them clearly with respect to other typical processes in strong interactions. It should also be noted that mixing may be at work worsening the situation.

\subsection{$\eta'$ decay}

\begin{figure}[tbp]
\begin{center}
\includegraphics[width=320pt]{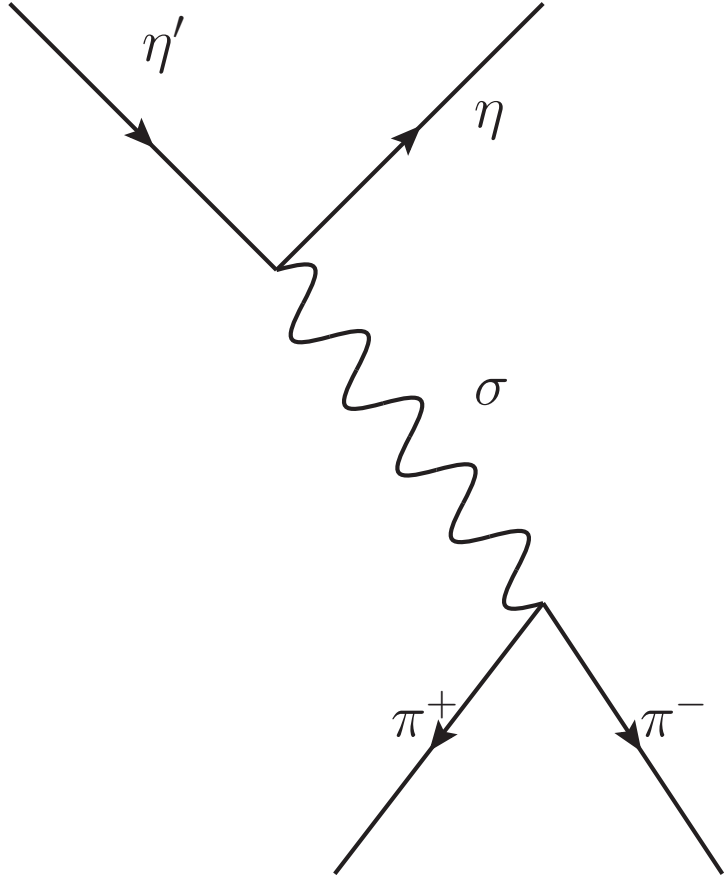}
\caption{\label{fig:fig2} $\eta'\rightarrow\eta+\pi^++\pi^-$ diagram.}
\end{center}
\end{figure}

\begin{figure}[tbp]
\begin{center}
\includegraphics[width=320pt]{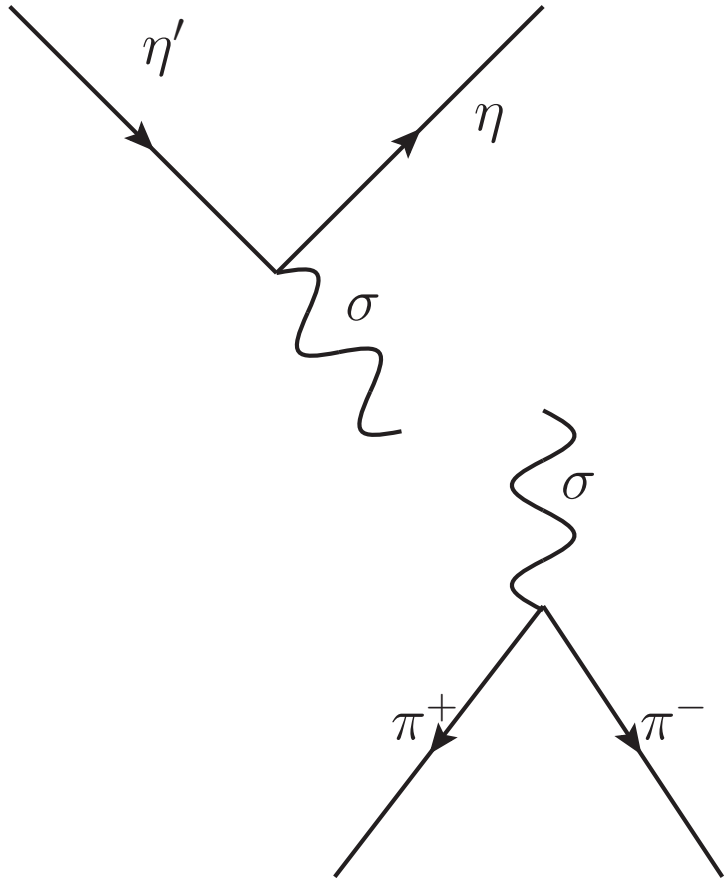}
\caption{\label{fig:fig3} $\eta'\rightarrow\eta+\sigma$, $\sigma\rightarrow\pi^++\pi^-$ diagrams.}
\end{center}
\end{figure}

Computation of the decay of $\eta'$ to $\eta\pi^+\pi^-$ is an interesting question in QCD. Chiral perturbation theory, at the leading order, is very far off the measured value \cite{bij}. So, it is interesting to see how low-energy limit of QCD recovers the amplitude. In our case, the diagram is given in fig.\ref{fig:fig2}
In this case there is a virtual $\sigma$ meson exchanged and this is a two-vertex process. 
In the following, using our approach, we exploit the possibility that $\eta'$ is emitting a $\sigma$ meson and we get the rate. Indeed, we can cut this diagram in two simpler ones in fig.\ref{fig:fig3}.
From this latter diagram we can consider $\eta'$ decay just as a production of $\sigma$ particles and we can immediately put down the amplitude for the production process $\eta'\rightarrow\eta+\sigma$ as
\begin{equation}
   |{\cal M}_{if}|^2=(N^2-1){G'}^{2}_{NJL} m_{\eta'}^2f_\sigma^2 f_\eta^2
\end{equation}
remembering that now this is a two-body decay. So, the width will be given by
\begin{equation}
   \Gamma_{\eta'} = \frac{1}{2\pi}{G'}^{2}_{NJL}m_{\eta'}f_\sigma^2 f_\eta^2
   \sqrt{\frac{m_{\eta'}^4+m_\eta^4+m_\sigma^4-2m_{\eta'}^2m_\eta^2
   -2m_{\eta'}^2m_\sigma^2-2m_\eta^2 m_\sigma^2}{m_{\eta'}^4}}.
\end{equation}
From eq.(\ref{eq:fsn}) we have
\begin{equation}
\label{eq:fs}
   f_\sigma=f_{S_0}=\sqrt{\frac{\sigma}{6\pi\alpha_s}}
   \frac{2\pi}{K(i)}\frac{e^{-\frac{\pi}{2}}}{1+e^{-\pi}}
\end{equation}
and we are in a position to give an estimation of the mass of the $\sigma$ resonance assuming that the threshold must be overcome for this process to be seen as is. Using the mass formula for the $\sigma$ we get the limit values $\sqrt{\sigma}=0.342\ GeV$ being $m_\sigma=0.410\ GeV$ that are quite near current estimations (e.g. see \cite{nar4}). Finally, taking $f_\eta\approx 0.019\ GeV$, very near $f_{\eta'}\approx 0.024\ GeV$ \cite{nv2}, and $m_{\eta'}=0.958\ GeV$ we have in the end $\Gamma\approx 91\ keV$ in agreement with the experimental value (see \cite{pdg}). This result is quite satisfactory and could give a proper account of the observed process. Indeed, with our understanding of this decay, we will have $\Gamma_{\eta'\pi\pi}=\Gamma_{\eta'}\Gamma_{\sigma}/(\Gamma_{\eta'}+\Gamma_{\sigma})\approx \Gamma_{\eta'}$ taking e.g. $\Gamma_\sigma/2=279^{+9}_{-12.5}\ MeV$ as given in \cite{lcc}. We can conclude that this process could be quite effective to extract the properties of the $\sigma$ resonance.

\section{\label{sec6} Conclusions}

We have given a consistent low-energy limit of QCD when a strong coupling approximation is considered. In this paper we have shown simple applications to work out the consequences of our approximations. This technique could be extended at higher orders but having a well-definite model at the leading order can already help to work out most of the phenomenology observed in current experiments.

\begin{acknowledgments}
I would like to thank Stephan Narison for helpful correspondence about this matter.
Feynman diagrams have been drawn using JaxoDraw \cite{db}. I would like to thank authors of this tool that makes so easy this work.
\end{acknowledgments}


\begin{thebibliography}{99}
\bibitem{gross} D. J. Gross and F. Wilczek, Phys. Rev. Lett. {\bf 30}, 1343 (1973). 
\bibitem{poli} H. D. Politzer, Phys. Rev. Lett. {\bf 30}, 1346 (1973).
\bibitem{wils} K. G. Wilson, Phys. Rev. D {\bf 10}, 2445 (1974).
\bibitem{wei} S. Weinberg, Physica {\bf A96}, 327 (1979). 
\bibitem{gleu} J. Gasser and H. Leutwyler, Ann. Phys. (N.Y.) {\bf 158}, 142 (1984).
\bibitem{sche} S. Scherer, Prog. Part. Nucl. Phys. {\bf 64}, 1 (2010).
\bibitem{na48} B. Bloch-Devaux, PoS(Confinement8)029 (2008). 
\bibitem{sre} M. Srednicki, {\sl Quantum Field Theory}, (Cambridge University Press, Cambridge, 2007). 
\bibitem{njl1} Y. Nambu and G. Jona-Lasinio, Phys. Rev. {\bf 122}, 345 (1961).
\bibitem{njl2} Y. Nambu and G. Jona-Lasinio, Phys. Rev. {\bf 124}, 246 (1961).
\bibitem{kle} S. P. Klevansky, Rev. Mod. Phys. {\bf 64}, 649 (1992).
\bibitem{rkl} K. Langfeld, C. Kettner and H. Reinhardt, Nucl. Phys. A {\bf 608}, 331 (1996).
\bibitem{fra0} M. Frasca, Int. J. Mod. Phys. E {\bf 18}, 693 (2009). 
\bibitem{ebe} D. Ebert, arXiv:hep-ph/9710511v1.
\bibitem{kle2} J. M\"uller and S. P. Klevansky, Phys. Rev. C {\bf 50}, 410 (1994).
\bibitem{vsz1} M. A. Shifman, A.I. Vainshtein, V. I. Zakharov, Nucl. Phys. {\bf B147}, 385 (1979). 
\bibitem{vsz2} M. A. Shifman, A.I. Vainshtein, V. I. Zakharov, Nucl. Phys. {\bf B147}, 448 (1979).
\bibitem{nar1} S. Narison, {QCD as a Theory of Hadrons}, (Cambridge University Press, Cambridge, 2004).
\bibitem{lcc} I. Caprini, G. Colangelo, Leutwyler H., Phys. Rev. Lett. 96, 132001 (2006).
\bibitem{pel} F. J. Yndurain, R. Garcia-Martin, J. R. Pelaez, Phys. Rev. D 76, 074034 (2007).
\bibitem{cre} V. Crede, C. A. Meyer, Prog. Part. Nucl. Phys. {\bf 63}, 74 (2009).
\bibitem{tg} T. Goldman, R. W. Haymaker, Phys. Rev. D {\bf 24}, 724 (1981).
\bibitem{rob} R. T. Cahill and C. D. Roberts, Phys. Rev. D {\bf 32}, 2419 (1985).
\bibitem{avs} R. Alkofer, L. von Smekal, Phys. Rept. {\bf 353}, 281 (2001).
\bibitem{cuc} A. Cucchieri, T. Mendes, PoS(LATTICE 2007)297 (2007).
\bibitem{ste} A. Sternbeck, L. von Smekal, D. B. Leinweber, A. G. Williams, PoS(LATTICE 2007)340 (2007).
\bibitem{vsz3} V. A. Novikov, M. A. Shifman, A. I. Vainshtein and V. I. Zakharov, Nucl. Phys. {\bf B229}, 381 (1983).
\bibitem{san} T. A. Ryttov, F. Sannino, Phys. Rev. D {\bf 78}, 065001 (2008).
\bibitem{boc} M. Bochicchio, PoS(EPS-HEP 2009)075 (2009).
\bibitem{fra1} M. Frasca, Phys. Rev. D {\bf 73}, 027701 (2006); Erratum-ibid., 049902 (2006).
\bibitem{fra11} M. Frasca, arXiv:0909.2428v2 [hep-th].
\bibitem{sus1} I. M. Suslov, arXiv:0911.1149v1 [hep-th].
\bibitem{sus2} I. M. Suslov, arXiv:0804.0368v3 [hep-ph].
\bibitem{fra12} M. Frasca, Phys. Lett. {\bf B670}, 73 (2008).
\bibitem{fra13} M. Frasca, Mod. Phys. Lett. A {\bf 24}, 2425 (2009).
\bibitem{fra14} M. Frasca, Int. J. Mod. Phys. D {\bf 15}, 1373 (2006).
\bibitem{smi} A. V. Smilga, {\sl Lectures on Quantum Chromodynamics}, (World Scientific, Singapore, 2001).
\bibitem{fra2} M. Frasca, Mod. Phys. Lett. A {\bf 22}, 1293 (2007).
\bibitem{fra3} M. Frasca, Int. J. Mod. Phys. A {\bf 23}, 299 (2008).
\bibitem{gr} I. S. Gradshteyn, I. M. Ryzhik, {\sl Table of Integrals, Series, and Products},
(Academic Press, 2000).
\bibitem{nv1} S. Narison, G. Veneziano, Int. J. Mod. Phys. A {\bf 4}, 2751 (1989).
\bibitem{nar2} G. Mennessier, S. Narison, W. Ochs, Phys. Lett. {\bf B665}, 205 (2008).
\bibitem{nar3} R. Kaminski, G. Mennessier, S. Narison, Phys. Lett. {\bf B680}, 148 (2009).
\bibitem{nar4} G. Mennessier, S. Narison, X.-G. Wang, arXiv:1002.1402v1 [hep-ph].
\bibitem{tep} B. Lucini, M. Teper, U. Wenger, JHEP 06, 012 (2004).
\bibitem{mor} Y. Chen, A. Alexandru, S.J. Dong, T. Draper, I. Horvath, F.X. Lee, K.F. Liu, N. Mathur, C. Morningstar, M. Peardon, S. Tamhankar, B.L. Young, J.B. Zhang, Phys. Rev. D {\bf 73}, 014516 (2006).
\bibitem{raf} R. Frigori, arXiv:0912.2871v1 [hep-lat].
\bibitem{cmn} C. McNeile, Nucl. Phys. B - Proc. Supp. {\bf 186}, 264 (2009).
\bibitem{bij} J. Bijnens, Acta Phys. Slov. {\bf 56}, 305 (2006).
\bibitem{nv2} S. Narison, G. M. Shore, G. Veneziano, Nucl. Phys. {\bf B433}, 209 (1995).
\bibitem{pdg} C. Amsler et al. (Particle Data Group), Physics Letters {\bf B667}, 1 (2008) and
2009 partial update for the 2010 edition.
\bibitem{db} D. Binosi, J. Collins, C. Kaufhold, L.Theussl, arXiv:0811.4113v1 [hep-ph].
\end{thebibliography}
\end{document}